# Automated Dyadic Data Recorder (ADDR) Framework and Analysis of Facial Cues in Deceptive Communication

T. SEN, University of Rochester
K. HASAN, University Rochester
Z. TEICHER, University Rochester
M. HOQUE, University Rochester

We developed an online framework that can automatically pair two crowd-sourced participants, prompt them to follow a research protocol, and record their audio and video on a remote server. The framework comprises two web applications: an Automatic Quality Gatekeeper for ensuring only high quality crowd-sourced participants are recruited for the study, and a Session Controller which directs participants to play a research protocol, such as an interrogation game. This framework was used to run a research study for analyzing facial expressions during honest and deceptive communication using a novel interrogation protocol. The protocol gathers two sets of nonverbal facial cues in participants: features expressed during questions relating to the interrogation topic and features expressed during control questions. The framework and protocol were used to gather 151 dyadic conversations (1.3 million video frames). Interrogators who were lied to expressed the smile-related lip corner puller cue more often than interrogators who were being told the truth, suggesting that facial cues from interrogators may be useful in evaluating the honesty of witnesses in some contexts. Overall, these results demonstrate that this framework is capable of gathering high quality data which can identify statistically significant results in a communication study.

## 1 INTRODUCTION

In 2009, Nicholas George, a traveler through Philadelphia National Airport, was pulled aside by Transportation Security Agency (TSA) behavior detection officers, specialists trained in reading nonverbal expressions indicative of deception and malicious intent [1]. Even though Nicholas was ultimately never charged with a crime, he was handcuffed and subjected to hours of detention by agents. After receiving micro expression training, TSA agents were still unsuccessful at identifying at least 16 travelers who were later found to be associated with terrorism [2]. Advancing our understanding of deception is a national priority to try to prevent the scenarios described above.

In the past, researchers have extensively studied the role of nonverbal behavior in light of recognizing deceptive intent during face-to-face interactions. Despite the generation of a body of new social science literature, most individuals are still slightly above chance in differentiating truth from lies [4]. Past studies have repeatedly come to the conclusion that humans, even expert interrogators, rarely do better than random chance in detecting deceit, with estimates not varying far from an accuracy rate of 54% [4,5]. What makes this problem seemingly intractable?

There are several factors that confound the detection of deception. Human expressions are considered spontaneous, multidimensional, and culturally sensitive. Using 43 muscles of our face, it has been estimated that humans can produce approximately 10,000 unique facial expressions at any given time [6]. Cultural nuances, gender, and age differences in facial expression add more to the complexity in the context of deception. Developing an automated system that can take all these variations into account would require a massive dataset with participants following a nontrivial protocol.

Furthermore, there are challenges to designing interaction protocols simulating real-life deception. A protocol needs to ensure that the participants are taking the task seriously and the collected data is likely to be spontaneous and natural. Most of the previous studies on deception involved bringing participants into a laboratory setting. However, a laboratory environment is unnatural for participants and is bound by logistical limitations due to infrastructure and resources, leading to limited sample sizes. For example, two of the largest deception video datasets ever collected involving deception obtained N=176 [7] and N=121 [8] dyads. Additionally, deception study data sets are usually not public. Another challenge for deception studies is having uncertain ground truths, in which the question of whether the participants are lying or telling the truth is never truly known. Due to the lack of large, good, public data sets on deception, using data driven technology to understand deception has been substantially limited.

In this paper, we developed the Automated Dyadic Data Recorder (ADDR) Framework for online dyadic conversation that provides researchers a large degree of automation in gathering deception data without bringing participants into a lab. This framework was used to run a novel interrogation protocol, in which paired participants were assigned either the role of a witness or an interrogator. The witness was directed by the automated system to either lie or tell the truth regarding presented evidence, while the interrogator was provided interactive instructions from the computer on how to question the witness. By leveraging crowd-sourcing and automation, our framework provides a large degree of automation in gathering dyadic audio-video data at anytime from virtually anywhere. Additionally, our framework is easily customizable to different research contexts.

Using this framework and interrogation protocol, we gathered N=151 dyads of data. To analyze the collected data and to form hypotheses regarding the nonverbal characteristics of deception, we leveraged two well-known psychological/communication theories: Duping Delight [9] and Interpersonal Deception Theory [10]. The theory of Duping Delight suggests that deceivers take pleasure in the act of deceiving and that pleasure may be manifested in the facial expressions of the deceiver. Interpersonal Deception Theory (IDT) focuses on the interaction between a message sender and a message receiver and suggests that the characteristics of the interaction can help indicate whether a communication is deceptive or truthful. Automated facial expression extraction was utilized to analyze facial expressions and develop features related to Duping Delight and IDT. Our study identified several nonverbal characteristics that differ between honest and dishonest communication. In summary, our study:

- presents a new, largely automated dyadic data recorder framework that gives ubiquitous access to crowd-sourced participants to conduct a directed online video interrogation and record their audio and video interaction;
- introduces a novel computer-mediated questioning protocol that captures nonverbal facial expression levels of participants during relevant questioning as well as baseline analytical, memory-based, and stressful questions;
- finds that although deceptive witnesses smiled slightly more often than their honest counterparts on average, the difference was not substantial (Cohen's $d < 0.10$);
- suggests that the facial expressions of an uninformed interrogator (i.e. one who does not know the ground truth of witness honesty) is useful in determining the honesty of a witness by showing that: (1) interrogators expressed higher levels of the smile-related lip corner puller (AU12) during relevant questioning when their witnesses were lying (Cohen's $d = 0.40$), and (2) interrogators returned a higher percentage of lying witness smiles during relevant questioning (Cohen's $d = 0.53$).

To the best of our knowledge, this paper is the first substantive experimental study using a highly-automated protocol for capturing dyadic video over the Internet to study deception. Analysis of the data generated insights on existing psychological theories of deception. Our protocol and the developed ADDR framework enables an experimenter to remotely collect behavioral data which otherwise may have been difficult to conduct in a laboratory setting. In addition, we identify many difficulties and limitations of this crowd-sourced approach and offer some practical solutions.

## 2 BACKGROUND

There are several challenges associated with gathering large and naturalistic data sets for the study of nonverbal behavior in deception. First, the deception protocol developed must ensure that the collected data is not only consistent, but also contains a high level of realism for the results to be useful. Second, because nonverbal behavior varies greatly between individuals, there is a need for personalized analysis, and development of features which are stable across individual nonverbal behavior differences. Third, if online automation and crowd-sourcing are to be leveraged for gathering large data sets, actions need to be taken to ensure that online video and worker quality does not suffer.

### 2.1 Consistency vs. Realism of the Protocol

Our desire to study deception in realistic environments competes with the need to gather data in consistent contexts. On one extreme, data can be gathered from real-life environments. For example, McQuaid et al. used video recordings of real-life television videos of both genuine and guilty individuals pleading for the return of their missing family members [11], and Perez-Rosas et al. used recordings from real-life criminal trial data [12]. While these studies demonstrate a high level of realism and high-stakes lies (and truths), they suffer from inconsistent video perspectives, non-uniform questioning of witnesses, and ultimately do not have ground truths known with absolute certainty. On the other extreme, researchers have achieved very high levels of consistency using fully computerized avatar interrogators. Derrick et al. developed a kiosk-based interrogator for lie detection [13], and Hoque et al. developed a responsive avatar for interviewing potential job candidates [14]. Fully automated avatar systems however appear to suffer from diminished realism and largely lack the ability to gather interactive, synchrony-based features of dyadic communication. Because of the inherent tradeoffs between consistency and realism it is unlikely that the "perfect data set" can ever be realistically gathered (Fig. 1).

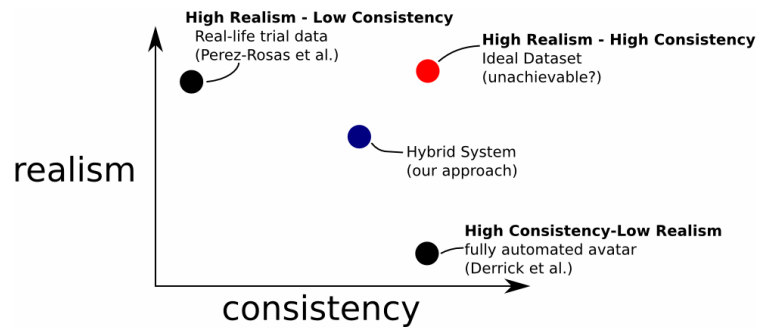

**Fig. 1. Consistency vs. Realism Tradeoff in Deception Protocols for Collecting Data.**

### 2.2 Prior Work in Nonverbal Indicators of Deception

Many theories of deceptive communication are based on the premise that a person's internal state of mind unconsciously and uncontrollably leaks out into the externally observable world where it can be detected. Since the early scientific investigations of this idea by Darwin in 1872, numerous other studies have researched nonverbal indicators of deception with limited success. In a study of animal and human expressions, Darwin postulated that particular facial muscles are difficult to repress and can reveal our true emotions [43]. Almost a century later, Paul Ekman popularized the theory that nonverbal cues are more indicative than verbal behavior in determining deception [15]. Ekman developed the notion of micro expressions, involuntary expressions lasting less than one fifth of a second that can be helpful in revealing true emotions [15]. Ekman utilized the notion of micro expressions to investigate the leakage of facial cues when people try to mask expressions concordant with lying [44, 45]. Other studies also found differences between genuine and fake emotions including happiness and disgust [46, 47] Nonverbal features related to cognitive load have been investigated under the premise that lying is often cognitively harder than recalling the truth [9]. Due to the high cognitive load, people may slow down their speaking rate, increase their pauses, and leak facial cues to show their true emotion while fabricating a lie. Characteristics associated with faked emotion were also investigated by ten Brinke et al., finding that people hesitate more and display emotional leakage when falsifying remorse [49].

Several studies analyzed hand gestures and head movements to detect deception. Lu, et al. showed that hand and head movements are useful to understand deceptive behavior [16]. Cohen, et al. found that deceptive people produced gestures which contradict the semantic information encoded in their speech [17]. Caso, et al. demonstrated that deceptive speakers use more metaphoric and rhythmic gestures [18].

Out of all the nonverbal communication features, smile is perhaps one of the most universal and commonly used expressions, and its relevance in deception has been investigated. The theory of Duping Delight asserts that liars experience pleasure in carrying out a deception [9]. While numerous references have talked about Duping Delight, few experimental studies have directly investigated the characteristics of smile with regards to deception. The largest published study identified directly related to Duping Delight (N=78 witnesses; 74,731 video frames; 41.5 minutes of video) is by ten Brinke, et. al, in which lower face expressions of happiness (coded using the authors' own manual coding system) were found to be displayed more often in deceivers [19]. A study of deception in three-year-olds (N=33), also found that smile was more strongly associated with deceivers [20]. Rosas et al indirectly investigated smile as one of several non-verbal modalities in building a multimodal deception detection classifier (N=121 with an average video length of 28 seconds) [12]. While their multimodal decision tree classifier was able to achieve an accuracy of 75%, their analysis does not identify any specific differences in smile between honest and dishonest communicators.

Instead of focusing solely on the message sender's behavior for a determination of truth, Interpersonal Deception Theory ("IDT") investigates characteristics of the interaction between a message sender and receiver for signs of deception [10]. A recent study investigated IDT's postulate that synchronized behavior is reduced during deception, and found that synchronized head nodding is greater in truthful communications [8].

## 2.3 Baselining Expression and Evoking Mental States

Most studies of nonverbal behavior look at absolute levels of a person's nonverbal expression without accounting for the individual differences or norms an individual may possess in their normal style of communicating. However, some questioning methodologies, such as the common polygraph Control Question Technique (CQT) [3, 21], take advantage of baselining, i.e. obtaining a measurement of an individual's nonverbal expression during questioning which is not relevant to the ultimate question of truth the interrogation wishes to determine. More specifically, the CQT involves asking a subject three types of questions: non-relevant questions, control questions, and relevant questions. The non-relevant questions are used to gauge a person's typical, normal behavior during communication. A control question is an uncomfortable, stress-inducing question, which is not related to the matter the person is being ultimately questioned for, but is used to gauge a person's behavior under stress. Finally, a relevant question involves the main question of guilt being investigated. A person's behavior is compared between these three types of questioning to help determine whether a person is being honest or deceptive. The Guilty Knowledge Test [22,42], used in criminal interrogations in Japan [23], attempts to evoke a specific type of nonverbal behavior in response to being presented evidence that only a guilty party would have knowledge of. Studies have used EEG, MRI, and eye tracking to measure subject's responses to guilty knowledge [24–27].

## 2.4 Leveraging the Internet and Automation

In addition to a good questioning protocol which can leverage baseline feature levels, a study needs to obtain a substantial number of data points in order to conduct meaningful analysis. In order to obtain a high level of statistical significance and to be able to apply machine learning techniques to data with more than a few input dimensions, larger data sets than have been contemporarily available are needed. Participants in video-based deception studies have been limited in size due to the limited availability of real-life data, and/or due to the cost of bringing participants into a laboratory setting. Past research using automation in non-video based deception studies highlights the potential that automation has to gather large amounts of data rapidly at a low cost. For example, a fully automated system for online dyadic text chat communication gathered over 3000 chat responses (half deceptive, half truthful) from participants [28]. Additionally, the emergence of several crowd-sourcing-based worker services provides a large pool of potential research subjects to help gather audio/video-based deception data faster and on a larger scale than is easily attainable in laboratory-based settings. For example, Manuvinakurike, et al. [29] designed an audio framework to collect spoken dialogue interactions between two remote human participants using crowd-sourcing. More specifically, they recorded the audio from a game where users had to guess correct image by exchanging information between them using their dialogue system.

Online video conferencing, as opposed to just chat or audio recording, introduces a whole set of computer system compatibility and video quality issues. Internet network conditions directly affect the resulting video resolution, frame rate, and person to person latency. For example, a study found that approximately 1 out of 5 Skype calls results in a loss of more than 1.5% data packets or suffer from end to end latencies of greater than 350 msec [30]. In addition, a participant's video stream is likely to suffer from poor webcam/microphone/computer quality, poor lighting, or background noise in his/her setting. Because of the many different potential ways in which online video quality can suffer, a framework for gathering online dyadic video data must place focus on maintaining video quality. Additionally, crowd-sourced workers are often associated poor quality [31]. Various methodologies have been investigated to obtain crowd-sourced worker quality, with widely varying levels of success [31,32]. Overall, in order to attain both high quality online research participants and high quality video conferencing, substantial effort must be made in developing an appropriate quality-assurance system.

## 3   FRAMEWORK DESIGN

In order to study nonverbal behavior associated with deception on a scalable level, we first developed a system for automatically gathering dyadic data. While generic options for online video calling and recording existed [33], no system existed for simultaneously directing participants to "play through" our desired interrogation questioning protocol while recording video at a high quality. By building a framework instead of a single system, we provide others the ability to rapidly develop new systems with a modified study protocol and/or context. In order to take advantage of the scalability of online services and the real-life setting of participants in his/her own home environment, we developed our framework as an Internet-based platform. In order to overcome the problem of inconsistent quality online workers and questionable quality video, our framework incorporates quality control as a fundamental element of the final system.

### 3.1   Technical Framework

*3.1.1 Overview.* The ADDR Software Framework ("the Framework") provides an efficient methodology and customizable software system to deploy a semi-automatic data gathering system for Internet-based dyadic communication studies. The Framework allows a researcher to rapidly tailor an ADDR system to run his/her particular research protocol. For example, we developed a system using the Framework to semi-autonomously direct crowd-sourced participants to follow the interrogation questioning protocol described in the next section and record the dyadic video data. With the ADDR Framework, researchers are largely freed from the burden of developing a web application from scratch and are further protected from the cost of discarding large portions of the collected data due to quality issues that are common with crowd-sourced workers.

The ADDR Framework comprises a videoconferencing online media server and a rudimentary video game state machine engine, which together are configured to provide a simple web browser interface for recruitment from a crowd-sourcing service. The primary components of the ADDR Framework are shown in Fig. 2 and include the Automated Quality Gatekeeper web application and the Session Controller web application. The Automated Quality Gatekeeper's primary purpose is to qualify potential participants by verifying that their computer system, network, and audio-video capture hardware is capable of gathering high quality data. The Automated Quality Gatekeeper additionally includes a researcher-customized Comprehension Test, used to verify whether a potential participant demonstrates sufficient understanding of any protocol-related directions provided to the potential participants during recruitment.  The Session Controller web application allows qualified participants to be linked over the internet and directs them through the steps specified in the Customized Protocol. Audio and video from each participant is recorded by the Session Controller as well as any participant responses to user interface questions presented by the Session Controller to the participants. The following sections discuss the ADDR framework in detail from a chronological perspective.

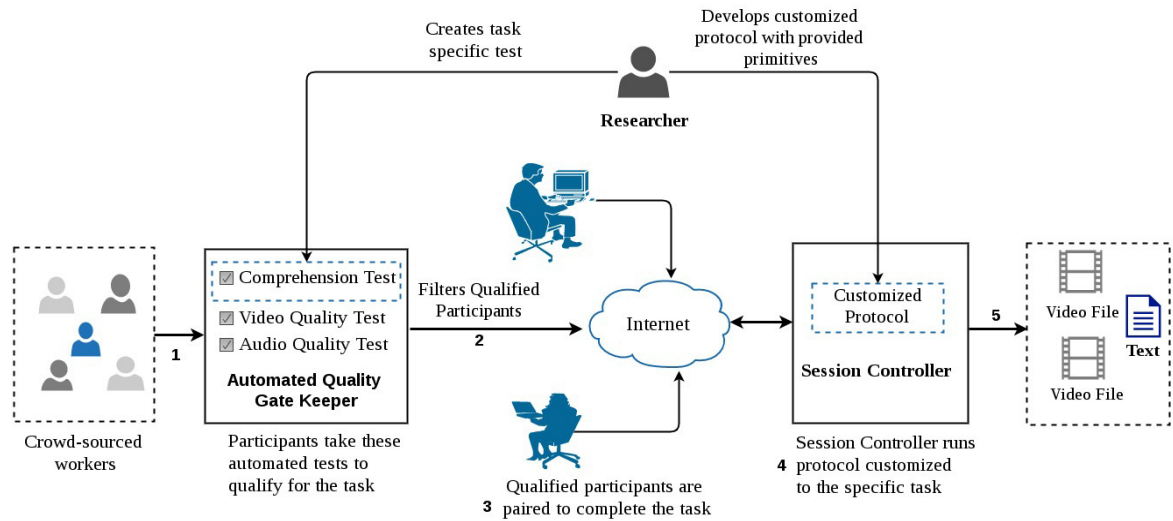

**Fig. 2. Overview of the ADDR Framework:** 1) crowd-sourced workers attempt to qualify as participants, 2) only workers who pass the Comprehension, Video Quality, and Audio Quality Tests are allowed to participate, 3) participants are paired and linked through a web application-based video conferencing session, 4) the Session Controller directs the participants to follow a Customized Research protocol (ex. our interrogation protocol), 5) upon completion of the protocol, recordings and UI data captured during the session are saved

*3.1.2 Automated Quality Gatekeeper.* The Automated Quality Gatekeeper ("Gatekeeper") web application's purpose is to automatically screen potential participants and determine if they have the necessary computer hardware, network bandwidth, and sufficient environmental conditions (such as lighting and noise free environment) to record high quality video. The Quality Gatekeeper also provides directions on how to play the protocol game, and obtains informed consent for study participation. [move: tests a participant's understanding of the research protocol directions provided]. The Quality Gatekeeper is an Apache Tomcat java web application built as a single page application (SPA) using the Kurento media server [34] and was hosted on a linode virtual linux server (2 GB ram, 1 CPU core, 30 GB SSD, bandwidth: 40 Gbps in, 1000 Mbps out). The Gatekeeper is based on the Kurento magic-mirror example source code. Upon first arriving at the Gatekeeper's URL, see Fig. 3 below, an informational page, asks the participant to ensure they have a webcam, broadband internet connection (2 Mbps upload and 2Mbps download), and compatible web browser (Firefox). While both the Firefox and Chrome web browsers were theoretically compatible with our system, because we experienced a higher number of video issues with Chrome, we limited participants to use Firefox to promote consistent system performance. Usage of the Firefox web browser is enforced using a java script userAgent check since we found many potential participants ignoring directions to use a particular web browser. As shown in Fig. 3, the Gatekeeper directs users to click on a start button. Upon first pressing the start button, the system checks the server to client to server lag. If which initiates a video loopback from the user's machine, to the web application located on a server, which is then looped back to the user's web browser. The Gatekeeper includes the OpenCv Facedetect plugin, which is used to detect whether a face is found with high confidence and appropriate size in the user's video stream. If such a face is detected, a green box is overlaid on the user's face together with an access code to continue. When user's click the "Next step" button shown in Fig. 3, they are prompted to enter the access code in order to continue. Additionally user's a provided a list of suggests on how they "fix" their system if either the video or the green box is not appearing. If a user's video does not appear in the local stream, they are encouraged to check their webcam functionality. If video does not appear in the server return stream, they are encouraged to check that they have provided permission to the browser to use the webcam and that other applications are not using the webcam. If the green box is not appeared they are directed to ensure they have good lighting of the face without a bright window in the background (this was a very common problem), and that their face takes up 1/2 to 3/4 of the video frame height.

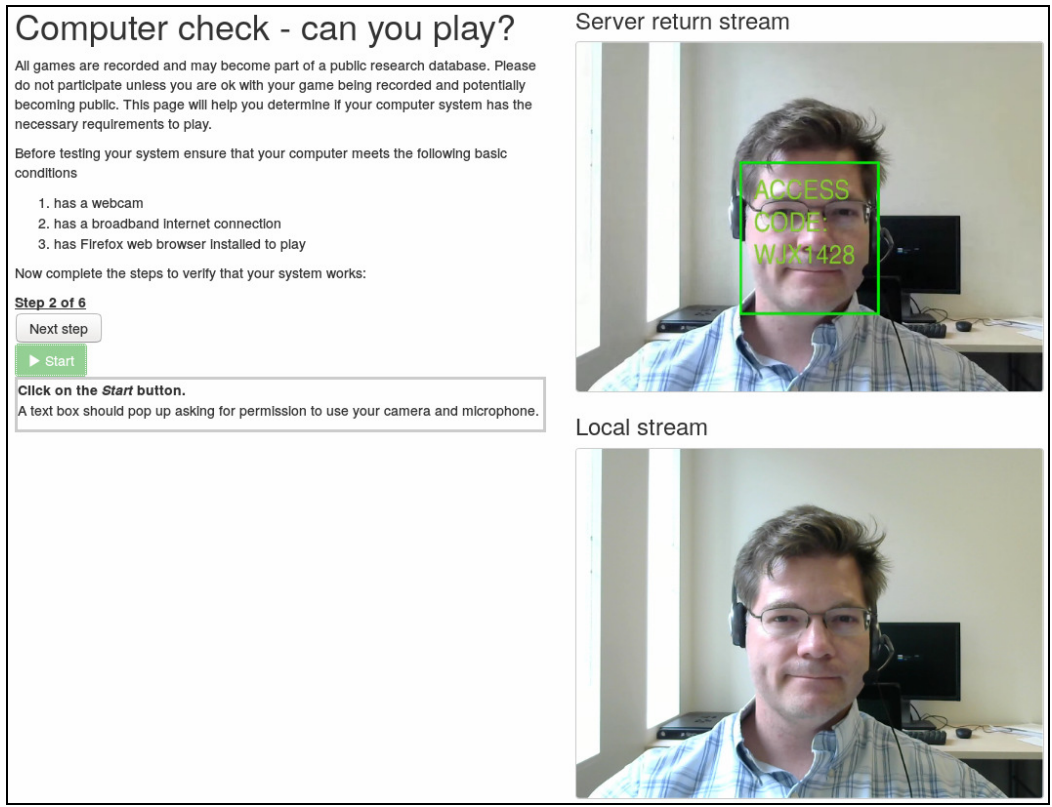

Fig. 3. Automated Gatekeeper User Interface

Participants are then directed to speak aloud to confirm that their audio is working and understandable. Upon pressing the "Next step" button again, the Gatekeeper provides a link to an instructional video detailing how the participants must follow the interrogation game protocol (i.e. that the witness will be shown an image for 30 sec, the computer will then tell the witness whether to tell the truth or lie about the image, the computer will then direct the interrogator what questions to ask the witness). The object of the game and the bonus payments are very explicitly mentioned in the instructional video (i.e. the witness's job is to always get the interrogator to believe them, regardless of whether they are lying or not; the interrogator's job is to correctly determine if the witness is lying.) Within the instructional video a second access code is provided. The Gatekeeper prompts users to enter this second access code in order to continue the qualification process. The Gatekeeper next conducts the Comprehension Test – a researcher customized test to demonstrate an understanding of the protocol. The framework provides a simple skeleton in which a researcher may easily specify a series of multiple choice or fill in the blank questions which are then presented to the user as a simple web page form. If a sufficient score is obtained on the Comprehension Test, users are presented a registration page in which they provide a username, password, and email address. Users are also provided a link to an informed consent document approved by our university's IRB. Before users can submit their registration, they are required to provide consent to participate in the study. Once registered, participants are emailed a link which they must click on in order to activate their registration. The email additionally provides them the URL for the Session Controller and informs them that they must schedule a game session.

*3.1.3 Crowd-Sourced Recruitment & Scheduling Mechanism.* Shown in Fig. 2, the Crowd-Sourced Recruitment & Scheduling Mechanism ("R&S") serves the primary purposes of (1) recruiting and directing potential research participants to visit the web URL of the Automated Quality Gatekeeper application in order to qualify for the study, and (2) schedule participants to play the protocol "game" with another participant. The R&S mechanism may be a formal crowd-sourcing platform, such as Amazon Mechanical Turk (AMT). Alternatively, crowd-sourced recruitment may be as simple as sending an email blast to potential participants, such as university students in a particular class. The particular crowd-sourcing service utilized is likely to affect several important factors including: scale (the number of potential participants that can be recruited in a given time); participant system quality (e.g. video camera and lighting quality, network bandwidth and lag, etc.); and participant quality (the likelihood that the potential participants will understand and follow directions, show up at their scheduled session time, and take the study seriously). While getting participants to visit the Gatekeeper URL to become qualified was not very difficult, scheduling participants to play a specific time. Our deployed instance of the ADDR framework was developed using both AMT recruitment and email based recruitment.

Recruitment with AMT. The basic unit of work or a job task on AMT is called a Human Intelligence Task or "HIT". The AMT crowd-sourcing site provides a marketplace where requestors creates a HIT describing a job task with a completion price and workers who are qualified may accept a HIT. An accepted HIT is then either returned for another worker to accept or is submitted when completed by the worker. A requester than may either accept the competed HIT, after which money is exchanged from the requestor's AMT account to the workers, or may reject the HIT if the requester is unsatisfied with the completed work. The AMT system contains a qualification system, in which workers may obtain qualifications specified either by AMT or requestors. For example, AMT specified qualifications include a workers country of residence, gender, age, as well as number of HITs completed by a worker and a workers submission acceptance rate. Requestor-specified qualifications include passing online tests created by the requestor. When creating HITs, a requestor may specify a number of qualifications which a worker must have before they are able to accept a HIT. We created a AMT qualification test in which a worker must provide the correct access codes which were provided during a participant's registration with the Gatekeeper. This ensured that only AMT workers who had successfully qualified with the Gatekeeper could accept our HITs. Each of HITs we created provided workers with a brief description of the interrogation game, specified a specific time in which they must play the game. While all workers could view our HIT job tasks, only workers who went to the Gatekeeper URL and registered would be able to obtain the AMT qualification necessary to accept our HITs. For example, we entitled a batch of HITs as "Play a 20 minute Interrogation Game Wednesday at 7pm for $10." The body of the HIT provided a simple description of the interrogation game and provided instruction for how a potential worker could qualify for the HIT by qualifying with our Gatekeeper. A qualified worker could accept the HIT, after which they were given 15 minutes to submit their HIT. The time for playing the game was always set to be several hours in the future and HITs were always configured to expire before the game time. Because our system is capable of supporting up to 4 concurrent games without suffering performance issues, each batch of HITs we created for a specific time included 8 HITs, such that up to 8 people could play concurrently (in four separate games). This multi-step scheduling process was necessary to address the substantial problems we had with AMT workers. These problems included, but are not limited to workers not showing up for their assigned time slot, workers not following directions, workers accepting HITs only to return them much later thus preventing other workers from signing up in the meantime.

Recruitment via email. Our implementation of email-blast-based recruitment involved having an a priori list of potential participants and made use of a shared online document (google docs) to act as a signup sheet among potential participants. Approval was obtained from our IRB to send a recruitment email computer science students enrolled at our university. The recruitment email provided a basic description of the interrogation game protocol and stated that students could earn $10-$20 for playing an online game involving lie detection. The email directed students to first see if they had compatible computer systems by registering at the Gatekeeper URL. Registered participants were invited to sign up on an online spreadsheet specifying online slots. The number of slots was limited in order to encourage students to sign up in the same time slot so that they are paired with another student.

*3.1.4 Session Controller.* The Session Controller web application allows registered participants to log into the system, and then establishes telecommunication links between pairs of participants which it immediately begins recording. The Session Controller is also implemented as an Apache Tomcat java web application built as a Single Web Page web application using the Kurento media server [34]. It was based on the Kurento one2one-call example source code. The Session Controller was run on a separate but similar capability server as the Gatekeeper to insure that sporadic loads of users registering with the Gatekeeper application do not cause problems with the Session Controller. The Session Controller interactively provides each participant directions and user interfaces as specified in the Customized Protocol and runs through the Protocol game states. The Session Controller also controls recording of participant audio video, provides participants with varied instructions depending upon the game state, and obtains interaction data from participants as laid out in the Protocol. An example user interface is shown in Fig. 4 below for a participant following our interrogation protocol. Note that the video display takes up the majority of the screen and shown in the video panel is the witness. Above the video frame the Session Controller provides directions and any questions specified according to the research protocol. The user interface is capable of providing buttons to the user interface which controls games state behavior. In the example user interface shown below, the interrogator gets the next question by clicking on the "Next Question" button. The time that the button was pressed is logged by the Session Controller such that the researcher may time stamp roughly when a particular question was asked. The video files, as well as the button presses are automatically recorded by the system for later retrieval by the researcher.

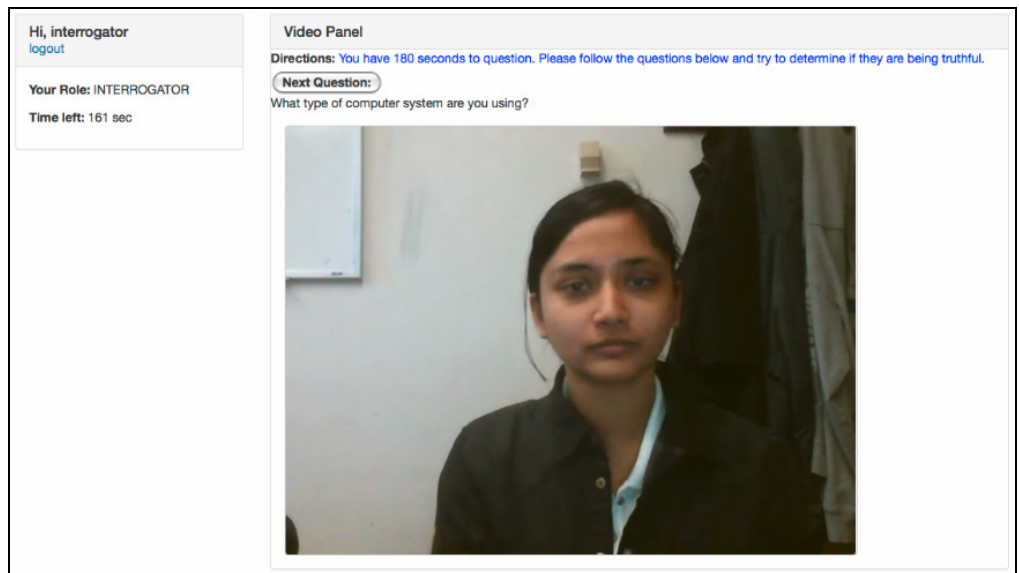

**Fig. 4. Example Interrogator User Interface**

## 3.2 Interrogation Questioning Protocol

Our experimental protocol was designed in light of the guiding principles, past studies, and psychological/communication theories described in the background section above. In summary, our protocol is a game between a witness and an interrogator. The witness is shown an evidence photograph and then directed to either be honest or lie to an interrogator. The interrogator is then guided to ask the witness a number of specific as well as self-generated questions.

Fig. 5 shows the different stages of our protocol. The protocol consists of the preliminary stages of Role Assignment, Evidence Review, and Witness Task Assignment (TRUTH/LIE), followed by three separate phases of questioning. The Role Assignment state begins after a video link is established between two participants and the participants indicate that they are ready to begin. In the Role Assignment State, the system assigns one participant to be a witness and the other participant to be an interrogator. The game then proceeds to the Evidence Review state. The Evidence Review state lasts for 30 seconds, during which the witness is shown an image on his/her user interface and instructed to memorize details of the image. The image is randomly selected from a set of ~50 photographs, an example of which is shown in step 2 of Fig. 5. The interrogator is instructed to wait during the Evidence Review state.

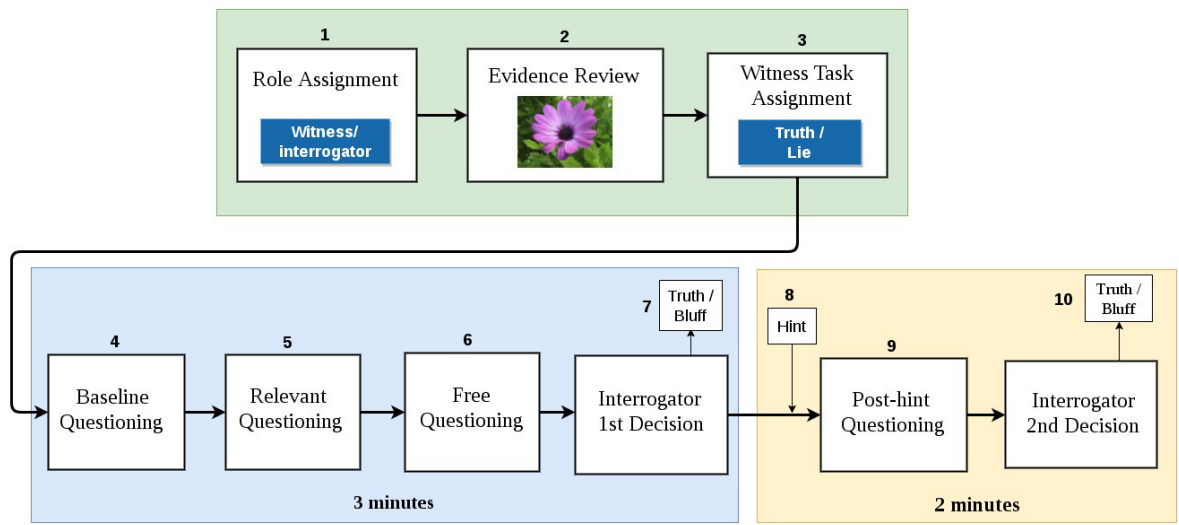

**Fig. 5. Stages of the Interrogation Protocol**. 1) Each of two participants is randomly assigned to be a witness or an interrogator, 2) witness reviews evidence image, 3) witness is randomly assigned to tell the TRUTH or to LIE, 4) interrogator is prompted to ask the witness Baseline Questions, 5) interrogator is prompted to ask the witness Relevant Questions, 6) interrogator is free to ask his/her own questions, 7) interrogator indicates whether he/she thinks witness is telling TRUTH or LIE about the evidence, 8) interrogator receives a text Hint, 9) interrogator is prompted to ask questions related to the Hint, 10) interrogator indicates his/her final decision regarding whether he/she believes the witness is telling a TRUTH or LIE regarding the evidence.

After 30 seconds have elapsed in the Evidence Review State, the protocol advances to the Witness Assignment State. In the Witness Assignment state, the witness is told whether they are to either tell the truth regarding the image they were just shown, or lie regarding the image and pretend that it was X instead (where X is a one or two word description of an object, such as ROCKET, or FLOWER POT). The protocol then advances to the Baseline Questioning state. In the Baseline Questioning state, the system instructs the witness to be honest regarding the Baseline Questions regardless of his/her assigned role of either telling the truth about the image or lying about the image. In the Baseline Questioning state, the system sequentially prompts the interrogator to ask the witness the Baseline Questions shown in Table 1. Each of the Baseline Questions were selected to evoke a certain mental state. The specific mental states we attempt to evoke and record include: a normal mental state, slight confusion, memory recall, analytic thought, and answering uncomfortable questions. For example, the question "What type of computer system are you using?" is ambiguous in that witnesses are likely to be confused as to whether they should report the computer type (i.e. desktop vs. laptop), the computer manufacturer, or the operating system. The interrogator is provided a NEXT QUESTION button on his/her user interface which causes the next question to be displayed. The system records the exact time that the button is pressed such that the approximate time that the question was asked can be automatically determined from the captured data. After all of the Baseline Questions have been asked, the game proceeds to the Relevant Questioning phase.

Table 1. **Baseline Questions & Evoked Mental State**

| Baseline Question | Evoked Mental State |
| --- | --- |
| Am I wearing glasses or not? | simple question normal mental state |
| What type of computer system are | ambiguous question slightly confused state |
| What color clothes did you wear | memory recall |
| What is 12 + 19? | analytic thought |
| Did you ever steal anything in your whole life and if so what was it? | discomfort |

Table 2. **Relevant Questions**

| Relevant Question |
| --- |
| What was your image? |
| Could you give me some more details about the image? |
| If there were something to count in the image, what would it be and what would be the count? |
| Were there any other objects in the image? |
| What are the colors in the image? |
| Please tell me about the background in the image. |
| Where do you think the photograph was taken? |
| Are parts of the object in the image man-made? |
| Continue asking your own questions. |

During the Relevant Questioning phase, the Interrogator is prompted by the system to ask broad and detailed questions regarding the image. Each of the relevant questions are listed in Table 2. The broad questions are designed to determine how much information the witness is willing to volunteer without prompting. The detailed questions are designed to evoke nonverbal expressions associated with different types of mental states in truthful and deceptive witnesses. In answering detailed questions, we anticipate liars to more often be in an analytical and relatively high cognitive load mental state as they formulate their deceptions. In contrast, we hypothesize that truthful witnesses will be more likely to express memory-based expressions as they recall details of their evidence images. The interrogator is provided three minutes of time to ask both the Baseline and Relevant Questions. Once the interrogator gets through the last Relevant Question, he/she is directed to continue asking his/her own questions (see Fig. 5, Free Questioning state). After the three minutes of time has elapsed from the beginning of the Baseline Questioning state, the system enters the Interrogator 1st Decision state. In this state, the video link is suspended and the interrogator is presented a dialog box asking the interrogator whether he/she believes that the witness is telling the truth or lying regarding the image. Once the Interrogator makes a decision, the system enters the Hint Questioning phase, in which the Interrogator is provided a text hint regarding the witness's evidence image, and is directed to ask the witness questions regarding the hint. The interrogator is provided two minutes to ask questions during the Hint Questioning phase. This phase was inspired by the Guilty Knowledge Test, and attempts to evoke different nonverbal expressions in witnesses who are lying vs. telling the truth. After two minutes in the Hint Questioning state has elapsed, the system again provides the interrogator a dialog box asking the interrogator whether he/she now believes the witness is being truthful or lying regarding the image.

In order to motivate the participants to take the game seriously and to raise the stakes of the witness's lies above a minimal level, payment to the participants was dependent upon their performance. More specifically, the witness and interrogator were paid a base payment of $10 for participating in the game, but could each earn an additional $5 or $10. The interrogator was paid an additional $5 for each correct decision he/she made regarding the witness's honesty (recall that the interrogator is asked twice by the system to make a decision regarding the witness.) The witness, regardless of whether he/she is telling the truth or lying, receives a $5 bonus for each interrogator decision believing that the witness is truthful. Thus, it is the witness's motivation is to always get the interrogator to believe him/her.

## 3.3 Research Questions

Our study aims to answer the following research questions:

**Q1: Is an automated crowd-sourced framework a viable strategy for automatically running and gathering data for dyadic online video communication experiments?** Are the quality issues related to crowd sourced workers using their own systems in their own homes surmountable?

**Q2: Do deceivers have different levels of nonverbal expression in online dyadic communication?** We specifically looked to examine smile-based facial cues and provide objective data to evaluate whether deceptive witnesses smile more often as posited by the theory of Duping Delight. In addition to examining findings related to smiling and deception, we also planned an exploratory study of differences in other facial expressions of witnesses and interrogators during different types of questioning. We were further curious of whether nonverbal interactions between a message sender and message receiver were different between honest and deceptive communications. Interactive Deception Theory [10] suggests that the interaction between message sender and receiver is indicative of deception. We hypothesized that the rate of shared smiles between witnesses and interrogators is an important measure of dyadic interaction and sought to identify whether this metric exhibits significant differences between deceptive and honest communication

## 3.4 Analysis Methods

In this section, we describe the feature extraction tools and the statistical methodology we used to evaluate our hypotheses.

*3.4.1 Facial Feature Extraction.* We used the OpenFace [37] open source software for automatic facial expression analysis. For each video frame (recorded at 15 fps), OpenFace estimates head pose, a number of facial landmark positions and a number of Facial Action Coding System (FACS) [15] outputs. OpenFace provides features based on the facial action coding system [38] which attempts to measure the movement of facial landmarks which roughly correspond to individual facial muscles (see Table 3). OpenFace's performance has been benchmarked on public manually coded databases [37]. We specifically we looked at Boolean valued AUs listed in Table 3 provided by OpenFace. The average of each of these features (i.e. the frequency of action unit expression) was calculated independently for both data recorded during the baseline questioning and relevant questioning phases, as well as for both truthful and deceptive witnesses and their interrogators. Upon visual inspection of plots of each of the features over time with regards to the associated video, it did not appear that the feature signals had high frequency noise. We thus did not apply smoothing/filtering.

Table 3. **Action Units extracted by the OpenFace analysis tool**

| Au Number | Description | Au Number | Description |
|---|---|---|---|
| AU01 | Inner Brow Raiser | AU14 | Dimpler |
| AU02 | Outer Brow Raiser | AU15 | Lip Corner Depressor |
| AU04 | Brow Lowerer | AU17 | Chin Raiser |
| AU05 | Upper Lid Raiser | AU20 | Lip stretcher |
| AU06 | Cheek Raiser | AU23 | Lip Tightener |
| AU07 | Lid Tightener | AU25 | Lips part |
| AU09 | Nose Wrinkler | AU26 | Jaw Drop |
| AU10 | Upper Lip Raiser | AU28 | Lip Suck |
| AU12 | Lip Corner Puller | AU45 | Blink |

*3.4.2 Percentage of Returned Smiles.* In addition to looking at average expression levels, we developed an algorithm to measure the percentage of returned smiles between interrogators and witnesses. Our algorithm for calculating the rate of returned smiles is directional. The percentage of interrogator smiles that are followed by a witness smile is calculated as shown in Fig. 5 and described in the following steps.

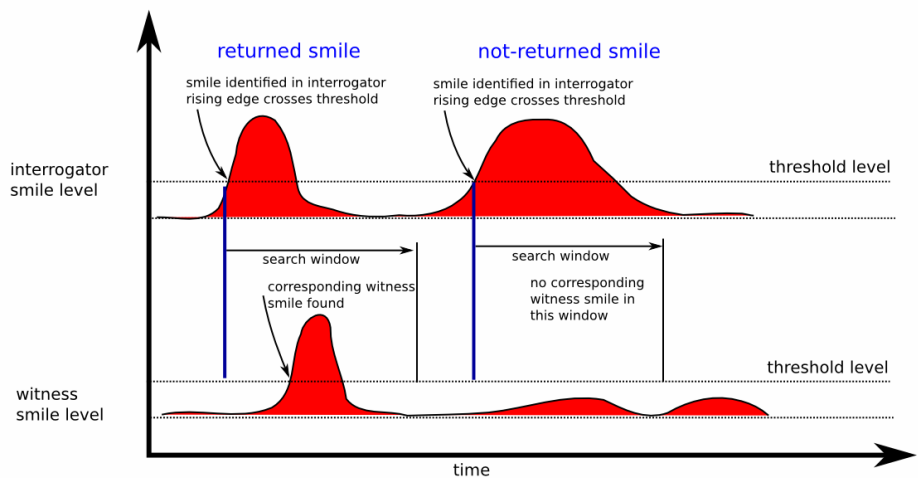

**Fig. 6. Example returned and not-returned smiles.**)

The process of calculating the percentage of returned smiles is shown in Fig. 5. First, if the interrogator's smile level crosses above the threshold level learned by the OpenFace, we mark it as the beginning of an interrogator smile. Then we search for a corresponding smile in the witness in the following 2.5 seconds. If a smile is found, it is counted as a returned smile. Below we describe how we calculate the percentage of returned smiles.

STEP 1: The algorithm calculates the total number of times that the interrogator smiles. A smile is defined as a positive rising edge of when the interrogator's OpenFace AU 12 Boolean output changes from a 0 to a 1.

STEP 2: For each interrogator smile rising edge, the algorithm searches the witness's smile data over a window spanning the 2.5 seconds after the interrogator's smile's rising edge. The algorithm searches for a "returned smile" in the witness's window. More specifically, the algorithm looks for a witness AU 12 output equal to 1 within the window. If a corresponding witness smile is found, a shared smile counter is incremented.

STEP 3: After scanning through all interrogator smile rising edges, the metric for the % returned smiles is calculated:

$$\text{Percentage of returned smiles} = \frac{\text{Number of interrogators smiles were returned}}{\text{Total number of interrogator smiles}}$$

Similarly, the % of witness smile's returned by the interrogator is calculated.

*3.4.3 Statistical Analysis.* We use the Student's unpaired t-test to test whether the mean smile frequency is different for the truthful and bluffing populations. The t-test assumes that the each of the distributions are normally distributed which may or may not be the case. Thus, we also used the Mann Whitney-Wilcoxon test (MWW) [39] in order to test whether the truthful and bluffing distributions have different medians (since the MWW test does not assume that the distributions are normally distributed). Cohen's d [40] is used to quantify the effect size, the difference between the truthful and bluffing frequencies in units of estimated standard deviation. Because the smile hypothesis is being tested in multiple ways (i.e. using AU 6 and AU 12, in personalized as well raw form, witnesses as well as interrogators; 2x2x2 = 8), a Bonferroni correction [41] of 8 times the p-values is discussed with regards to the stated p-values in the results and discussion.

## 4 RESULTS

## 4.1 Data Quality and Cleaning

The number of dyads who completed the game is N=398. This number includes both data that was gathered before implementation of the Automated Quality Gatekeeper and after its implementation. (before = 140, after = 258). The number of usable dyads before and after the Automated Quality Gatekeeper is 24 and 127 respectively. Thus, before the AQG was implemented 12% of the data were usable, and after its implementation, 73% were usable. The usability of data was dictated by a variety of reasons, most removals due to poor quality AV recording (ex. video approaching 1 FPS, video with substantial gaps in recording, lighting quality too poor to detect any faces). Other failures include individuals who: did not follow or understand directions, witnesses who did not notice the evidence image, participants who were eating during the session so as to disrupt face detection, participants who were moving out of the camera field of view, participants who had the camera oriented so poorly as to not detect faces, and participants with loud background noise. In summary, we obtained N=151 usable dyads for facial analysis (TRUTH=75, LIE=76).

## 4.2 Smile and Duping Delight

Shown in Table 4 below are the frequency levels of the AU 6 (cheek raiser) and AU 12 (lip corner puller) as extracted by the OpenFace automatic facial expression analysis tool. In the "Normalized" column, the "raw" data refers to a person's expression levels during the relevant questioning phase. The "personalized" row represents expression levels by subtracting an individual's baseline expression levels. Note that the interrogators do not have a "personalized" row since they were not asked baseline questions. As shown in Table 4, bluffing witnesses show a greater raw AU 6 frequency (0.273) compared to the truthful witnesses (0.247). However, t-test and MWW p-values > 0.05 indicate such differences are not statistically significant at a 95% confidence level. Similarly, bluffing witnesses showed higher raw AU 12 frequency (0.303) compared to the truthful witnesses (0.282). Again, the t-test and MWW p-values indicated that this difference is not statistically significant at a 95% confidence level. The personalized AU 6 and AU 12 metrics also showed higher levels in bluffing witnesses (-0.107, -0.162) as compared to truthful witnesses (-0.132, -0.165). Note that the personalized Truth and Bluff "frequencies" can be negative since they represent the smile frequency during baseline questioning subtracted from the smile frequency during relevant questioning. While these findings are in the direction of supporting the theory of Duping Delight, the low p-values and low Cohen's d values suggest that the effects predicted by Duping Delight can not be confirmed by our study.

While the witnesses did not exhibit significant differences in their smile frequencies, the interrogators showed substantial differences in their AU 12 frequencies. Interrogators paired with bluffing witnesses showed a higher AU 12 frequency (0.356) compared to interrogators paired with truthful witnesses (0.239). The uncorrected t-test and MWW test p-values are 0.017 and 0.005 respectively. Applying a Bonferroni correction of 8 brings these p-values to 0.136 and 0.04. Thus, the MWW test shows that the higher median AU 12 expression frequency for interrogators paired with bluffing witnesses is statistically significant at a confidence level of 95%. The Cohen's d effect size calculation indicates that the two distributions have an estimated difference of 0.397 standard deviations. It is important to note that the interrogators did not know the ground truth of whether the witness was bluffing or being truthful. The accuracy of interrogators in correctly judging the witnesses (prior to receiving the hint) was 57% overall (48% for interrogators paired with bluffing witnesses and 66% for interrogators paired with honest witnesses).

Table 4. **Comparison of Duping Delight Related Features Between Truthful and Lying Witnesses**
(Note: Bonferroni correction is not applied to table data. Please see text for discussion of Bonferroni correction.)

| Role | Normalization | Feature | Truthful group frequency | Bluffing group frequency | t-test | MWW | Cohen's d |
|---|---|---|---|---|---|---|---|
| witnesses | raw | AU06 | 0.247 | 0.273 | 0.551 | 0.373 | -0.099 |
| | personalized | AU06 | -0.132 | -0.107 | 0.494 | 0.227 | -0.113 |
| | raw | AU12 | 0.282 | 0.303 | 0.64 | 0.382 | -0.077 |
| | personalized | AU12 | -0.165 | -0.162 | 0.93 | 0.325 | -0.014 |
| interrogators | raw | AU06 | 0.242 | 0.283 | 0.357 | 0.13 | -0.152 |
| | raw | AU12 | 0.239 | 0.356 | 0.017 | 0.005 | -0.397 |

## 4.3 Smile Synchrony

The results of applying the returned smile algorithm to the data are shown in table 5 below. The percentage of witness smiles that follow interrogator smiles is higher in the truthful witness group (0.518 vs. 0.445 respectively). However, the t-test and MWW test p-values indicate that the difference is not statistically significant. The percentage of interrogator smiles that follow witness smiles is lower in the truthful group (0.397 vs. 0.580). This difference had p-values of 0.002 (0.008 with a Bonferroni correction factor of 4 due to using two tests with two different synchrony measures) and an effect size of over one half a standard deviation (d=0.530).

Table 5. **Percentage of returned smiles by witnesses and interrogators**

| Smile Order | Truthful group | Bluffing group | t-test | MWW | Cohen's d |
|---|---|---|---|---|---|
| Interrogator · Witness | 0.518 | 0.445 | 0.203 | 0.106 | -0.213 |
| Witness · Interrogator | 0.397 | 0.580 | **0.002** | **0.002** | 0.530 |

## 4.4 Exploratory Study of Facial Features

The results of the exploratory study of the 15 action units listed in table x other than AU 6 and AU 12 are shown in Table 5 below. Only action units showing a difference with an uncorrected p-value less than 0.05 are listed in the table. The only witness action unit showing a difference at this level is the personalized AU 15 (lip corner depressor) feature. Fig. x below shows an example expression of lip corner depressor by a bluffing witness. The truthful witnesses showed a lower level of personalized lip corner depressor frequency than bluffing witnesses (-0.044 and 0.031 respectively). The effect size of this difference is notable at d = 0.48, nearly approaching one-half standard deviation. Also, raw AU 1 (inner brow raiser) frequency was substantially higher in interrogators paired with truthful witnesses (0.199) compared to bluffing witnesses (0.160). Because the high number of tests performed (18 action units, two participants, two normalizations, two tests = 144 combinations) we are expected to find a p-values < 0.05 even if when no true difference exists. With application of a Bonferroni correction factor of 144, we clearly do not demonstrate statistical significance with either of these action unit findings. However, blind application of a Bonferroni correction of 144 is likely to introduce type 1 errors (rejecting differences when one indeed does exist.) It is reasonable to consider each of the separate 36 action units being investigated is a separate hypothesis and that the appropriate Bonferroni correction to apply is 4 (two tests: t-test and MWW; and two normalizations: raw and personalized). We recognize the contentious nature of how statistical tests should be corrected and merely provide all the data to the readers to draw their own conclusions.

Table 6. **Action Units showing a substantial difference in frequency of expression between honest and deceptive communication.** (Note: Bonferroni correction is not applied to table data. Please see text for discussion of Bonferroni correction.)

| Role | Normalization | Feature | Truthful group frequency | Bluffing group frequency | t-test | MWW | Cohen's d |
|---|---|---|---|---|---|---|---|
| witnesses | personalized | AU 15 | -0.044 | 0.031 | 0.004 | 0.005 | 0.481 |
| interrogators | raw | AU 1 | 0.199 | 0.160 | 0.085 | 0.049 | -0.285 |

## 5 DISCUSSION

### 5.1 Do deceivers have different levels of nonverbal expression?

The results presented in section 4.2 identified that the interrogators' AU 12, lip corner puller expression frequency was significantly higher when the witness was lying. The interrogators did not know the ground truth of whether the witness was lying or not and were correct in predicting whether the witness was lying only 57% of the time. Due to the low interrogator accuracy, it is unlikely that the differences in AU 12 expression are a result of the Interrogator knowing a witness is lying and finding their story amusing. Instead, it is likely that there are differences in deceptive witnesses' behavior that are causing the interrogators to smile more often. It is likely that deceptive witnesses are trying harder to get the interrogators to like them through various behavior, such as humor, hoping that an interrogator will feel more guilty of accusing the witness of being liar. This notion is supported by the returned smile statistics which show that interrogators are substantially more likely to return a lying witness's smile (35.6% vs. 23.9% for lying and honest witnesses respectively). Why would interrogators smile more in response to lying witnesses more than honest witnesses? Perhaps it is because lying witness smiles are tied to a conscious effort by the witness to influence the interrogator.

The notion that witnesses are modulating their behavior as a function of whether they are lying or not is grounded in IDT, which posits that messages senders modulate their behavior based on whether they are being honest or not. While past research has found that the level of synchrony goes down based on the veracity of the message sender, we found that the direction of synchrony is important. While the interrogator's smile behavior was more synchronized with dishonest witnesses (58% vs. 39.7% for dishonest and honest witnesses respectively), deceptive witnesses' behavior was less synchronized with the interrogators (44.5% vs. 51.8% for dishonest and honest witnesses respectively). Thus, we found that receiver following sender mimicry goes up with deception while sender following receiver mimicry goes down with deception.

In investigating the 16 other action units extracted by OpenFace, we were surprised that only one action unit, AU 15, lip corner depressor, showed any difference in frequency between lying and honest witnesses. However, the significance of this finding must be considered with caution. As mentioned in section 4.4, applying a Bonferroni correction factor of 144 renders the p-value of the AU 15 differences non-significant. However, each of the AU's is looking at a different measure, and in such cases, researchers have suggested it inappropriate to apply such Bonferroni correction. Ad hoc review of our data identified that lip corner depressor in several instances showed up when deceivers were showing faked, over-exaggerated expressions of trying to remember. This notion appears also in Bartlett et. al study of expressions of faked pain which identified that faked of expressions of pain were most readily identified by exaggerated open-mouth expressions [50]. This finding suggests that future studies investigate a priori hypotheses with regards to lip corner depressor expression in deceptive and honest communication.

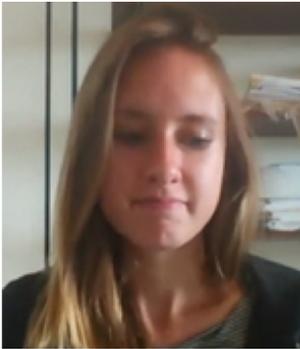

**Fig. 7. Bluffing participant pretending to try to remember**.

### 5.2  Is an automated crowd-sourced framework a viable strategy for automatically running and gathering data for dyadic communication experiments?

After gathering 151 high quality dyadic AV videos of crowd-sourced participants following our protocol and showing that this data demonstrates statistically significant communication findings, we conclude that the ADDR framework is indeed a viable system for automatically gathering data for dyadic communication experiments. However, the question of whether the ADDR framework is efficient must be carefully considered.

Our study also found that when personalized behavior is taken into account, some features, such as lip corner depressor become more helpful in predicting deception. We were surprised that out of all the personalized facial features we examined, we found a moderate effect size in only one feature: the lip corner depressor. We expected that while taking into account each individual's baseline expression levels, we would see stronger differences between deceptive and truthful witnesses. One possible explanation of this is that deceivers are conscious of the fact that they are going to be deceptive even when they may be honestly answering baseline questions. In explaining the increased levels of smile in deceivers as a potential greater social reward, we note that deceivers are conscious of that greater reward during both the baseline and relevant questioning. Therefore deceptive witnesses may still be experiencing duping delight even though they are not actively lying during baseline questioning.

Though our overall usable data rate (i.e. 151/398) is not good, it is not representative of our final system. The Automated Quality Gatekeeper was only added after experiencing substantial quality issues regarding both the video and worker quality. After all features of the Automated Quality Gatekeeper were added to the framework, usable data quality went up from 12% high quality dyads 73% high quality dyads (in which a high quality dyad is defined as having 82% face-trackable frames).

We plan to modify our research protocol to study other contexts of deception. In addition, our future analyses will study verbal and language cues involving in deception as well as temporal features associated with deception. Moreover, we working to make our framework more robust and secure, such that it can be deployed to gather data in numerous environments in the wild. With participant's permission, it is not unrealistic to apply our framework to gather data in conjunction with police interrogations, airport or border patrol screening, interviewing job applicants, and even online dating.

Our study demonstrates that the ADDR Framework can be an effective way to gather data useful in identifying nonverbal features associated with deception. Our study of deceptive communication found multiple results with statistical significance which validates the utility of our framework.

## 6   CONCLUSION

In this paper, we demonstrate that the ADDR framework is an effective system for gathering dyadic communication data through validation with a study of deceptive communication. The framework provides the ability to ubiquitously collect interrogation data from online crowd-sourced workers, which overcomes one of the major limitations in deception data collection. Moreover, we developed a novel interrogation protocol that makes use of questions which are analytical, memory-based, and stressful in order to study nonverbal expressions associated with different mental states. Inspired by the psychological theories of Duping Delight and IDT, we introduced several hypotheses about deceptive behavior and studied the statistical significance of these hypotheses. Our results showed that while deceptive witnesses tend to smile more the level of difference was not statistically significant. The study however did identify statistically significant differences in both the interrogator smile, finding that interrogators paired with deceptive witnesses smiled more. The study also identified that the percentage of witness smiles returned by the interrogator is higher when the witness is lying. These findings provide indication that our framework is capable of gathering dyadic data of a quality high enough to establish statistical significance in investigated hypotheses. Moreover, our protocol allows us to capture personalized features: measurements of an individual's deviations from his/her baseline nonverbal behavior. An analysis of personalized features found that positive deviation in lip corner depression above baseline expression levels is greater in deceptive witnesses. We hope that application of the ADDR framework will help researchers to gather larger data sets than was previously achievable.

### ACKNOWLEDGMENTS
This work was supported by Grant 66653-NS with the Army Research Office (ARO). The authors would like to thank professor Randall Nelson for his insight in developing the protocol and Cindy Ryan for her help in coding the ADDR protocol.